\documentclass[a4paper,amsmath,amssymb,aps,prl,twocolumn,longbibliography,english,superscriptaddress]{revtex4-1}
\usepackage{amssymb}
\usepackage{graphicx}
\usepackage{dcolumn}
\usepackage{bm}
\usepackage{amsmath}
\usepackage{subfigure}
\usepackage{float}
\usepackage{color}
\usepackage[colorlinks=true,citecolor=blue]{hyperref}

\begin{document}


\title{Microwave photo-association of fine-structure-induced Rydberg $(n+2)D_{5/2}nF_{J}$ macro-dimer molecules of cesium}

\author{Jingxu Bai}
\affiliation{State Key Laboratory of Quantum Optics and Quantum Optics Devices, Institute of Laser Spectroscopy, Shanxi University, Taiyuan 030006, China}

\author{Yuechun Jiao}
\email{ycjiao@sxu.edu.cn}
\affiliation{State Key Laboratory of Quantum Optics and Quantum Optics Devices, Institute of Laser Spectroscopy, Shanxi University, Taiyuan 030006, China}
\affiliation{Collaborative Innovation Center of Extreme Optics, Shanxi University, Taiyuan 030006, China}

\author{Rong Song}
\affiliation{State Key Laboratory of Quantum Optics and Quantum Optics Devices, Institute of Laser Spectroscopy, Shanxi University, Taiyuan 030006, China}

\author{Georg Raithel}%
\affiliation{Department of Physics, University of Michigan, Ann Arbor, Michigan 48109-1120, USA}

\author{Suotang Jia}%
\affiliation{State Key Laboratory of Quantum Optics and Quantum Optics Devices, Institute of Laser Spectroscopy, Shanxi University, Taiyuan 030006, China}
\affiliation{Collaborative Innovation Center of Extreme Optics, Shanxi University, Taiyuan 030006, China}

\author{Jianming Zhao}%
\email{zhaojm@sxu.edu.cn}
\affiliation{State Key Laboratory of Quantum Optics and Quantum Optics Devices, Institute of Laser Spectroscopy, Shanxi University, Taiyuan 030006, China}
\affiliation{Collaborative Innovation Center of Extreme Optics, Shanxi University, Taiyuan 030006, China}

\date{\today}

\begin{abstract}
Long-range $(n+2)D_{5/2} \, nF_J$ Rydberg macro-dimers are observed in an ultracold cesium Rydberg gas for $39\leq n\leq48$. Strong dipolar ``flip'' ($\langle D_{5/2} F_{5/2} \vert \hat{V}_{dd} \vert  F_{5/2} D_{5/2} \rangle$,
$\langle D_{5/2} F_{7/2} \vert \hat{V}_{dd} \vert  F_{7/2} D_{5/2} \rangle$)
and ``cross'' ($\langle D_{5/2} F_{7/2} \vert \hat{V}_{dd} \vert  F_{5/2} D_{5/2} \rangle$) couplings
lead to bound, fine-structure-mixed $(n+2)D_{5/2}nF_J$ macro-dimers at energies between the $F_J$ fine-structure levels. The $DF$ macro-dimers are measured by microwave photo-association from optically prepared
$[(n+2)D_{5/2}]_2$ Rydberg pair states. Calculated adiabatic potential curves are used to elucidate the underlying physics and to model the $DF$ macro-dimer spectra, with good overall agreement.
Microwave photo-association allows Franck-Condon tuning, which we have studied by varying the detuning of a Rydberg-atom excitation laser. Further, in Stark spectroscopy we have measured  molecular DC electric polarizabilities that are considerably larger than those of the atomic states. The large molecular polarizabilities may be caused by high-$\ell$ mixing. The observed linewidths of the Stark-shifted molecular lines provide initial evidence for intra-molecular induced-dipole-dipole interaction.
\end{abstract}


\maketitle

Ultracold gases in the $\mu$K-regime have opened a new avenue to the
investigation of interacting many-body systems. While interactions between ultracold ground-state atoms generally remain fairly weak, interactions involving Rydberg atoms (atoms with principal quantum numbers $n \gtrsim$10) 
are readily observable in typical cold atomic gases. 
Molecules in which a ground-state atom is located inside the Rydberg atom~\cite{Greene2000,Hamilton2002,Khuskivadze2002,Bendkowsky,Anderson2014,Krupp2014,Booth2015,Thomas2016,Sa2015,Bai2020} have attracted considerable attention due to their permanent electric dipole moments~\cite{Booth2015,Thomas2016}.
Rydberg-atom-ion molecules, which were predicted and observed~\cite{Duspayev2021,Dei2021,Zuber2022,Zou2023}, are bound by electric multipolar interactions, as are molecules consisting of two Rydberg atoms. The latter, known as Rydberg macro-dimers, were predicted in~\cite{Boisseau2002} and observed, for instance, in ~\cite{Overstreet2009,Deiglmayr2014,Saßmannshausen2016,Han2018,Hollerith2019,Hollerith2022,Hollerith2023}. Rydberg-pair interactions, which include the dipole-dipole ($\thicksim n^4/R^3$) and the van-der-Waals ($\thicksim n^{11}/R^6$)~\cite{Gallagher} interactions ($R$ is the internuclear distance), give rise to long-range adiabatic potential curves. Some of these have wells that support meta-stable Rydberg-pair macro-dimer molecules.

Deiglmayr {\sl{et al.}} have prepared Cs macro-dimer molecules near $nS \, n'F$ and $[nP]_2$ asymptotes for 22$\leq n \leq 32$~\cite{Deiglmayr2014}, and near $nP \, (n+1)S$ for $n$=43 and 44~\cite{Saßmannshausen2016}, where molecules are bound through long-range dipolar interaction.
Han {\sl{et al.}}~\cite{Han2018} have reported on $[62D]_2$ macrodimer molecules
prepared via a two-color, two-photon photo-association scheme, and have modeled their results using molecular-potential computations~\cite{Han2019}. Hollerith {\sl{et al.}}~\cite{Hollerith2019,Hollerith2022,Hollerith2023} directly observed Rydberg macrodimers with bond lengths up to 0.7~$\mu$m in optical lattices using an ion microscope, manipulated the molecular arrangement, and proposed applications in quantum computing.

\begin{figure}[htbp]
\begin{center}
\includegraphics[width=0.48\textwidth]{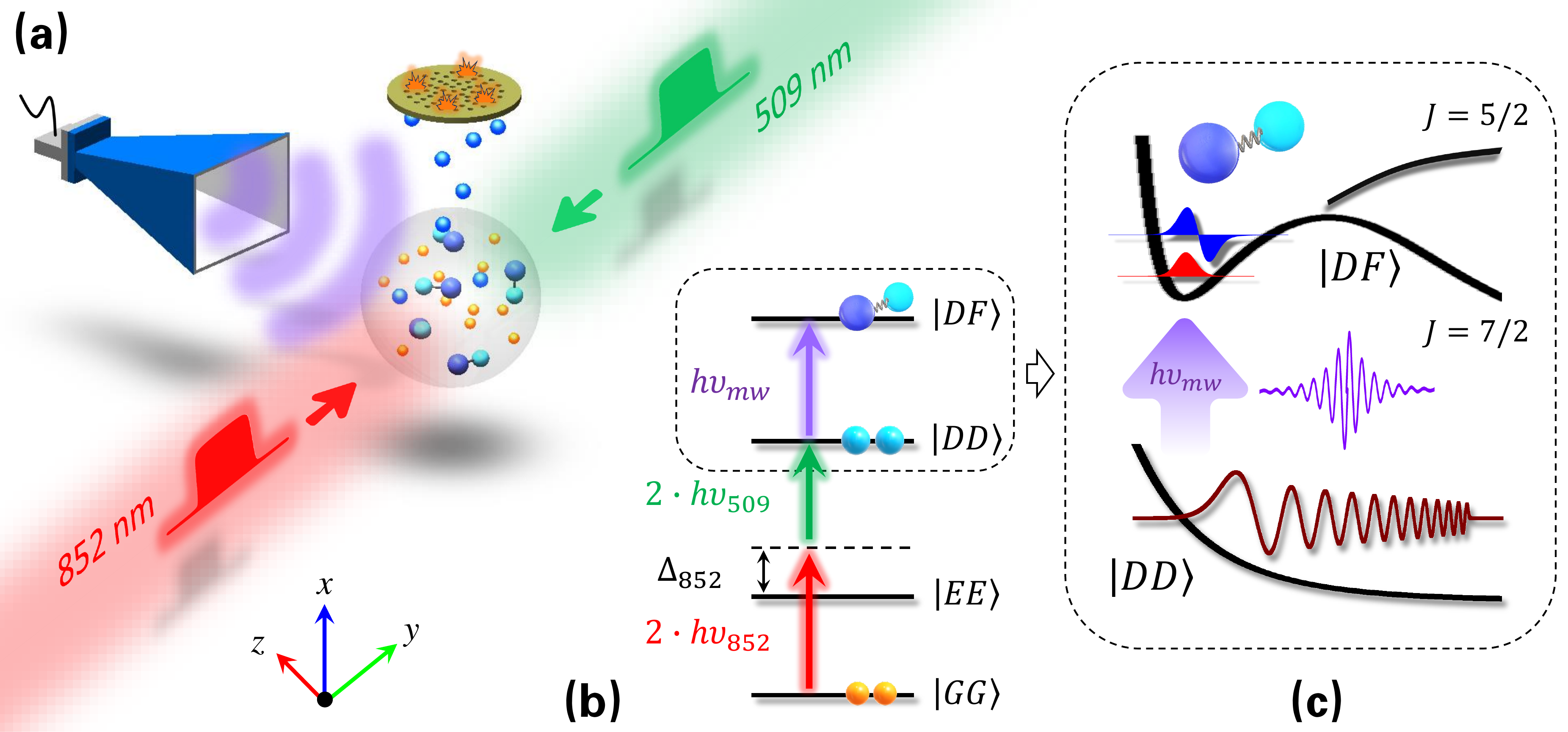}
\end{center}
\caption{(a) Sketch of the experimental setup. Rydberg macro-dimer molecules are produced by laser excitation of $D$-state Rydberg atoms and microwave photo-association.
(b) Level diagram. A Rydberg pair, $|DD\rangle = [(n+2)D_{5/2}]_2$, is laser-excited from $|GG\rangle =[6S_{1/2}]_2$ by 852-nm and 509-nm lasers (852-nm-laser detuning $\Delta_{852}/2\pi =360$~MHz).
A scanned microwave field excites $nF_{J}$ atoms or photo-associates $\vert DF \rangle= (n+2)D_{5/2} \, nF_{J}$ FS-mixed macro-dimers, which are detected as described in the text.
(c) Visualization of microwave photo-association with incident and bound vibrational wave-functions on respective potential curves. 
}
\label{Fig1}
\end{figure}

In this letter, we report on the observation of fine-structure-mixed (FS-mixed)
$(n+2)D_{5/2} \, nF_J$ Rydberg macro-dimer molecules bound by dipolar interactions, the longest-range multipole coupling between neutral atoms. 
Instead of using the common method of laser photo-association, here we prepare the $(n+2)D_{5/2} \, nF_J$ macro-dimers by microwave photo-association. The microwave field 
drives a transition from
an optically-prepared $[(n+2)D_{5/2}]_2$ Rydberg-pair state, which is relatively weakly-interacting, into a more strongly-interacting $(n+2)D_{5/2} \, nF_J$ macro-dimer state. We provide spectroscopic evidence for FS-mixed $DF$ macro-dimer molecules and explain their formation mechanism. 
We further address Franck-Condon tuning afforded by microwave photo-association spectroscopy, and investigate the Stark effect and the linewidths of $DF$ macro-dimer signals in weak electric fields.   


The experiment is performed in a Cs magneto-optical trap with a
temperature $\sim$~100~$\mu$K and peak density $\sim$~$10^{10}$~cm$^{-3}$. The atom temperature is decreased to $\sim$40~$\mu$K after 3~ms of molasses cooling. After switching off the molasses, a $(n+2)D_{5/2}$ Rydberg state is populated by coincident and counter-propagating 852-nm and 509-nm laser pulses of 1-$\mu$s duration, as shown in Fig.~\ref{Fig1}~(a). Rydberg-atom pairs are then photo-associated into $(n+2)D_{5/2} \, nF_J$ macro-dimers by a 20-$\mu$s microwave pulse.
An electric-field ramp is then applied for state-selective field ionization~\cite{Gallagher}. Time-resolved detection of the liberated ions with a microchannel plate detector and a boxcar integrator then allows selective detection of $nF$-type Rydberg atoms and molecular constituents. Prior to the experiment, we employ Stark and Zeeman spectroscopy to reduce stray electric and magnetic fields to below 2~mV/cm and 5~mG, respectively.
More experimental details are given in~\cite{bai2023}. 

\begin{figure}[htbp]
\begin{center}
\includegraphics[width=0.48\textwidth]{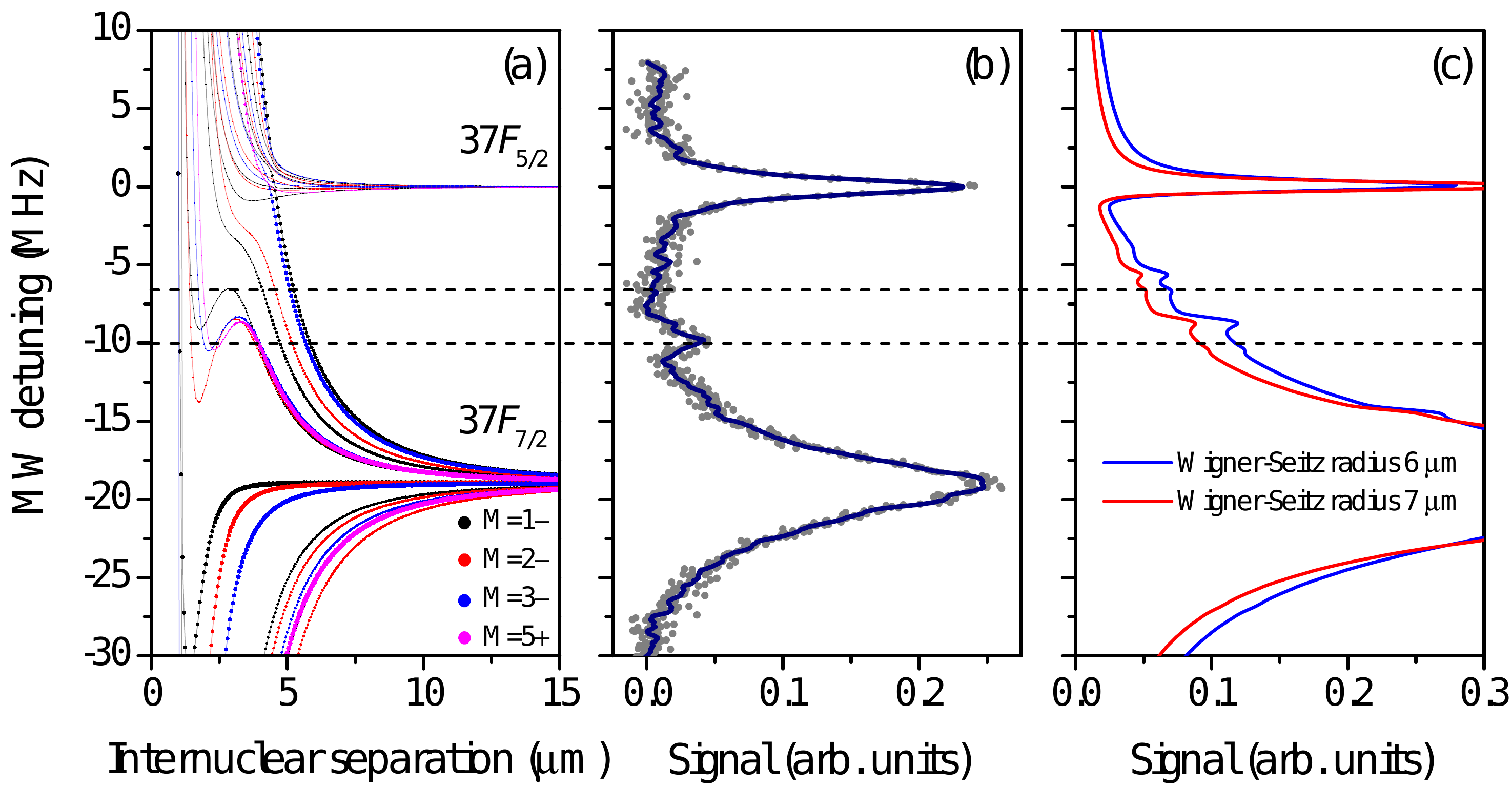}
\end{center}
\caption{
(a) Calculated potential energy curves vs internuclear separation $R$ for $39D_{5/2} \, 37F_{J}$ pairs with the indicated values for $M$ and parity ($-/+$ means odd/even). Symbol areas are proportional to calculated microwave transition rates from the launch state $[39D_{5/2}]_2$. 
(b) Experimental microwave spectrum, averaged over 50 measurements.
In (a), between $37F_{7/2}$ and $37F_{5/2}$, we find molecular potential wells that correspond with signals measured in (b).
These signatures reflect FS-mixed $39 D_{5/2} \, 37 F_{J}$ macro-dimers.
Dashed horizontal lines across (a) and (b) mark the detunings of prominent potential maxima and minima.
(c) Simulated microwave spectra for two Wigner-Seitz radii, showing good qualitative agreement with (b).
}
\label{Fig2}
\end{figure}

In Fig.~\ref{Fig2}~(a) we show calculated adiabatic potentials for Cs Rydberg pair states in $39D_{5/2} \, 37F_{J}$ for the indicated values of $M$, the conserved sum of the projections of the total electronic angular momenta onto the internuclear axis,
and parity. For details of the calculations see Supplemental Material~\cite{supp}. 
The detuning is defined to be zero at the $39D_{5/2} \, 37F_{5/2}$ asymptote. 
In Fig.~\ref{Fig2}~(b), we show the measured photo-association spectrum versus microwave detuning.
The narrow and wide features in (b) near zero detuning 
and $\approx - 20$~MHz show the 
$39D_{5/2}\to 37F_{5/2}$ and $39D_{5/2}\to 37F_{7/2}$ atomic resonances, which are broadened due to dipolar Rydberg-pair interactions. About halfway in-between the atomic resonances the spectrum in Fig.~\ref{Fig2}~(b) exhibits a narrow signal that coincides with Rydberg-dimer potential wells seen in Fig.~\ref{Fig2}~(a), as indicated with the dashed lines.

At large distances ($\gtrsim 8~\mu$m in Fig.~\ref{Fig2}), the dipole-dipole interactions are dominant and lead to asymptotic potentials $\propto C_3/R^3$. The $C_3$-coefficients are much larger for $(n+2)D_{5/2} \, nF_{7/2}$ than for $(n+2)D_{5/2} \, nF_{5/2}$ due to differences in angular matrix elements. Averaged over all $M$ and parities, the magnitudes of the $C_3$-coefficients average to $\sim 1.3$~GHz$\mu m^3$ and  $\sim 0.07$~GHz$\mu m^3$ for the $37F_{7/2}$ and $37F_{5/2}$ resonances, respectively.  
Also, the respective 96 and 72 long-range potentials are fairly evenly distributed and symmetric about the asymptotic energies. As a result, the interaction-broadened atomic $37F_J$-lines in Fig.~\ref{Fig2}~(b) are quite symmetric, with the $F_{7/2}$-line being wider than the $nF_{5/2}$-line by a factor of about 5.

The potential wells in Fig.~\ref{Fig2}~(a) correlate with signal maxima in the experimental data in Fig.~\ref{Fig2}~(b). To explain their origin, we first note that  the strongest interactions are 
the resonant ``flip'' interactions between (symmetrized) $\vert 39D_{5/2}, 37F_{7/2} \rangle$ pair states, which have dipole-dipole matrix elements 
$V_{7/2} := \langle D F_{7/2} \vert \hat{V}_{dd} \vert  F_{7/2} D \rangle$ and 
cause the aforementioned massive broadening of the atomic $F_{7/2}$-line. The  matrix elements of the $\vert 39D_{5/2}, 37F_{5/2} \rangle$ flip interactions, $V_{5/2} := \langle DF_{5/2} \vert \hat{V}_{dd} \vert  F_{5/2}D \rangle$, are only about $0.05$-times as strong, resulting in a less-broadened atomic $F_{5/2}$-line. With decreasing $R$, numerous $39D_{5/2} \, 37F_{7/2}$-potentials push upwards into comparatively weakly-shifting $39D_{5/2} \, 37F_{5/2}$-potentials and intersect, generating a series of anti-crossings with dipolar ``cross''-coupling strengths on the order of $\sqrt{V_{7/2} V_{5/2}}$, amounting to about a quarter of the atomic FS splitting. With all relevant couplings being dipolar and hence relatively strong, the potential wells that result for several $M$'s and parities are well-distinct and robust. The vibrational states in the wells are likely stable against non-adiabatic decay, as suggested by our experimental observations of sharp $DF$ macro-dimer features. For further argument, we note that, due to the proximity of the atomic FS states, the FS-mixed $DF$ wells are ``squeezed'' inside a spectral range that is only a few tens of MHz wide (for $n \sim 40$). The proximity of the atomic FS states is instrumental in pushing the radial positions of the mixed-FS $DF$ macro-dimer wells far out along $R$. There, intersections with other (potentially unaccounted-for) potentials are unlikely, or, if there are any, they are likely too weakly coupled to cause fast non-adiabatic decay.

To prove the FS-mixed character of the observed macro-dimers, we have computed $s(R) = \langle J - \ell \rangle (R)$ of the adiabatic electronic states on 
several potential curves in Fig.~\ref{Fig2}~(a) that have wells. 
Since $s = \pm 1/2$ for $J=\ell + 1/2$ and $J=\ell - 1/2$, $s$ is a measure for FS mixing.
At large $R$ outside the wells it is $s(R) \approx 1/2$, which is expected for $39D_{5/2}\, 37F_{7/2}$.
Inside the wells $s(R)$ drops to about $0.1$, indicating a FS-mixed $39D_{5/2} \, 37F_J$ macro-dimer. We have also find that mixing with $\ell \ge 4$-states is not a significant factor in the formation of the wells.

To model the experimental spectra, we have computed  microwave excitation rates from the optically-prepared launch state $[(n+2)D_{5/2}]_2$ into the adiabatic pair states $39D_{5/2} 37F_J$ on the potentials in Fig.~\ref{Fig2}~(a). Typical results are shown in Fig.~\ref{Fig2}~(c) for two indicated values of the Wigner-Seitz radius. For details of the calculation see Supplemental Material~\cite{supp}.
The model spectrum reproduces the essential features measured in Fig.~\ref{Fig2}~(b), except that the measured features attributed to the $DF$-molecules appear to be down-shifted by 1 to 2~MHz relative to the computed ones.
The difference is attributed to the positive, $\lesssim 2$~MHz, van-der-Waals level shifts of the launch pair, which result in likewise negative shifts of the measured molecular peaks. It is thus seen that energy shifts on the free potential in Fig.~\ref{Fig1}~(c), and Franck-Condon overlaps between the free and bound vibrational states, play a role in microwave photo-association spectroscopy of Rydberg macro-dimers.

\begin{figure}[htbp]
\begin{center}
\includegraphics[width=0.48\textwidth]{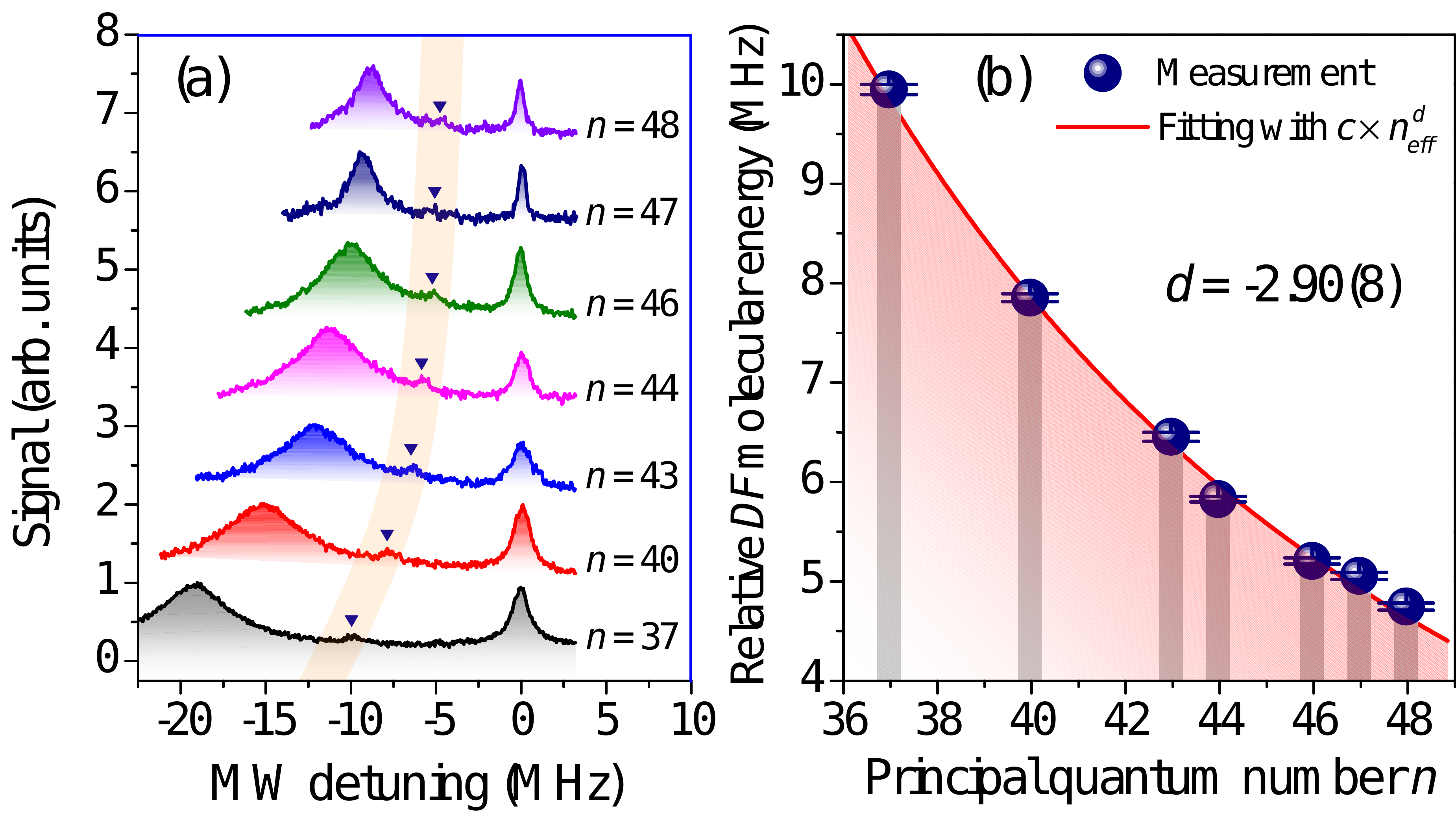}
\end{center}
\caption{(a) Microwave spectra, showing interaction-broadened atomic $nF_J$ lines (tall peaks on the left and right) and $(n+2)D_{5/2} \, nF_J$ macro-dimer signals (highlighted peaks in the middle)  for $37\leq n\leq48$. 
(b) Measurements (symbols) of the energies of the macro-dimer peaks relative to the $nF_{5/2}$ atomic lines
as a function of $n$. The solid red line shows an allometric fit with fit parameters as indicated.
}\label{Fig3}
\end{figure}

From microwave spectra analogous to Fig.~\ref{Fig2}~(b) we have extracted the energy of the most-prominent molecular $DF$ peaks relative to the $nF_{5/2}$ atomic lines for $37\leq n\leq48$. The magnitudes of the relative line positions, extracted from the spectra displayed in Fig.~\ref{Fig3}~(a) and plotted against $n$ in Fig.~\ref{Fig3}~(b), closely follow an allometric fit of the form $y=c \, n_{eff}^{d}$. Our fit result $d = -2.90(8)$ means that the line positions track closely in proportion with the FS splitting.
Macro-dimer energy scalings near $n^{-3}$ have been observed elsewhere~\cite{Samboy2011-1,Samboy2011-2,Han2018}. Here, the data and scalings observed in Fig.~\ref{Fig3} show that the FS-mixed $DF$ macro-dimers are generic and exist over a substantial range in $n$. The molecular potentials and the widths of the $DF$ molecular peaks in Figs.~\ref{Fig2} and~\ref{Fig3} further show that the depths of the molecular-potential wells ({\sl{i.e.}} the molecular binding energies) are about 25\% of the relative line positions shown in Fig.~\ref{Fig3}(b).

Findings above suggest that the energy of the $DD$ launch pair state on its repulsive interaction potential may be tuned to maximize the Franck-Condon overlap for the microwave transitions. To study this possibility, we have varied the detuning of the 509-nm laser to the blue side of atomic $nD_{5/2}$ resonances during the launch-state preparation.
For $n=48$, we have observed a net improvement of the molecular signal at about $200$~kHz blue detuning. The efficacy of this type of Franck-Condon tuning is hampered by the width of the initial kinetic-energy distribution of the atoms, which translates into a frequency uncertainty on the order of 1~MHz on the incident channel. Also, the blue-detuning leads to a general decrease of launch-state Rydberg-atom density, which has a negative effect on the net macro-dimer microwave excitation probability. Hence, our observation of a moderate molecular-signal enhancement when blue-detuning the Rydberg excitation laser is encouraging. Improvements of Franck-Condon tuning
may be achieved by reducing the atom temperature and by preparing the launch-state $nD_{5/2}$ atoms in a Rydberg-atom  optical lattice~\cite{Malinovsky2020,Cardman2023,Ravon2023}.

\begin{figure}[htbp]
\begin{center}
\includegraphics[width=0.48\textwidth]{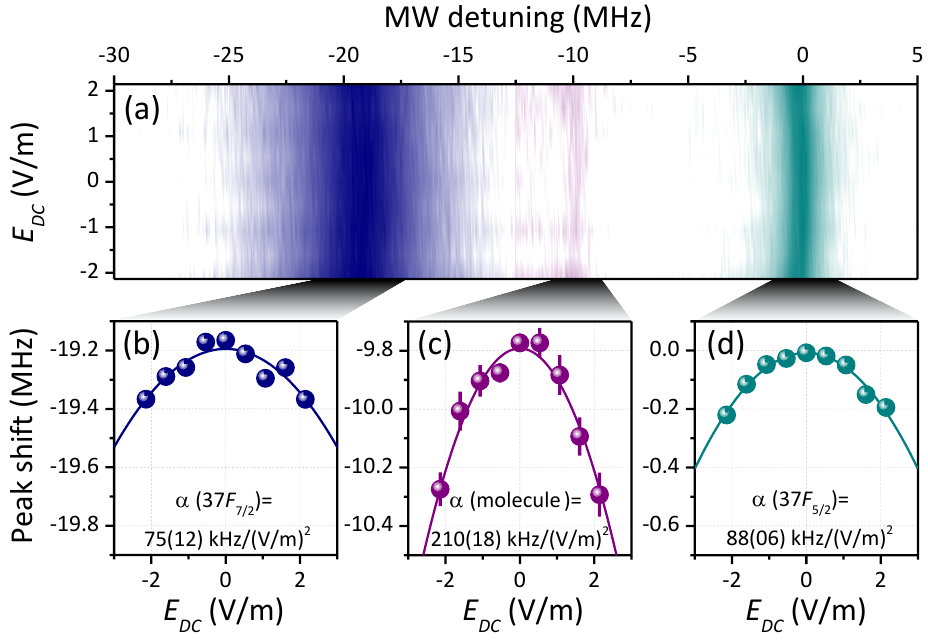}
\end{center}
\caption{
(a) Color-scale plot of microwave spectra near $37F_J$ vs DC electric field, $E_{DC}$, and microwave detuning. Symbols in the bottom panels show Stark shifts of the atomic line  $37F_{7/2}$ (b), the $39D_{5/2} \, 37F_J$ Rydberg macro-dimer (c), and the atomic line $37F_{5/2}$ (d), respectively. 
The solid lines show best-fit functions of the type $W= - 1/2\alpha E_{DC}^2 +\beta$, with fitted polarizabilities $\alpha$ shown in the panels.
}\label{Fig4}
\end{figure}

Next we measure the Stark effect of the $DF$ Rydberg macro-dimers in a weak applied static electric field, $E_{DC}$.
In Fig.~\ref{Fig4}(a), we show a contour plot of spectra versus $E_{DC}$ and microwave detuning. In Fig.~\ref{Fig4}(b-d), we present Stark shifts of the atomic and molecular peaks, obtained from Lorentzian fits.
Both atomic and molecular lines are red-shifted due to quadratic Stark effect, with fitted polarizabilities $\alpha$($37F_{5/2}$)=88(6)~kHz/(V/cm)$^2$, $\alpha$($37F_{7/2}$)=75(12)~kHz/(V/cm)$^2$ and $\alpha_M := \alpha$($39D_{5/2} \, 37F_J$)=210(18)~kHz/(V/cm)$^2$.
The polarizability $\alpha_M$ of the molecular state is found to be about 2.7 times larger than that of the atomic states. 

Calculated atomic polarizabilities (see Supplemental Material~\cite{supp}) yield an isotropic average of the differential polarizability of $\approx 85$~kHz/(V/m)$^2$ for the atomic lines in Fig.~\ref{Fig4}, which agrees with the measurements within uncertainties. For the isotropic average of the molecular polarizabilities, our calculation (see Supplemental Material~\cite{supp}) essentially yields the sum of the atomic polarizabilities, which is about a factor of 2.7 smaller than the measured $\alpha_M$. We believe that the calculation may fail to account for minute admixtures from highly polarizable $\ell \ge 4$ states. Increasing our basis sets to include atomic states with $\ell$ up to 15, there is evidence that the macro-dimer states pick up increasing (but still too small) high-$\ell$ character. 
It remains to be seen whether even larger basis sets will yield molecular polarizabilities in agreement with the measured $\alpha_M$. Also, the validity of our utilized perturbative model for molecular polarization (see Supplemental Material~\cite{supp}) could be tested.

\begin{figure}[htbp]
\begin{center}
\includegraphics[width=0.35\textwidth]{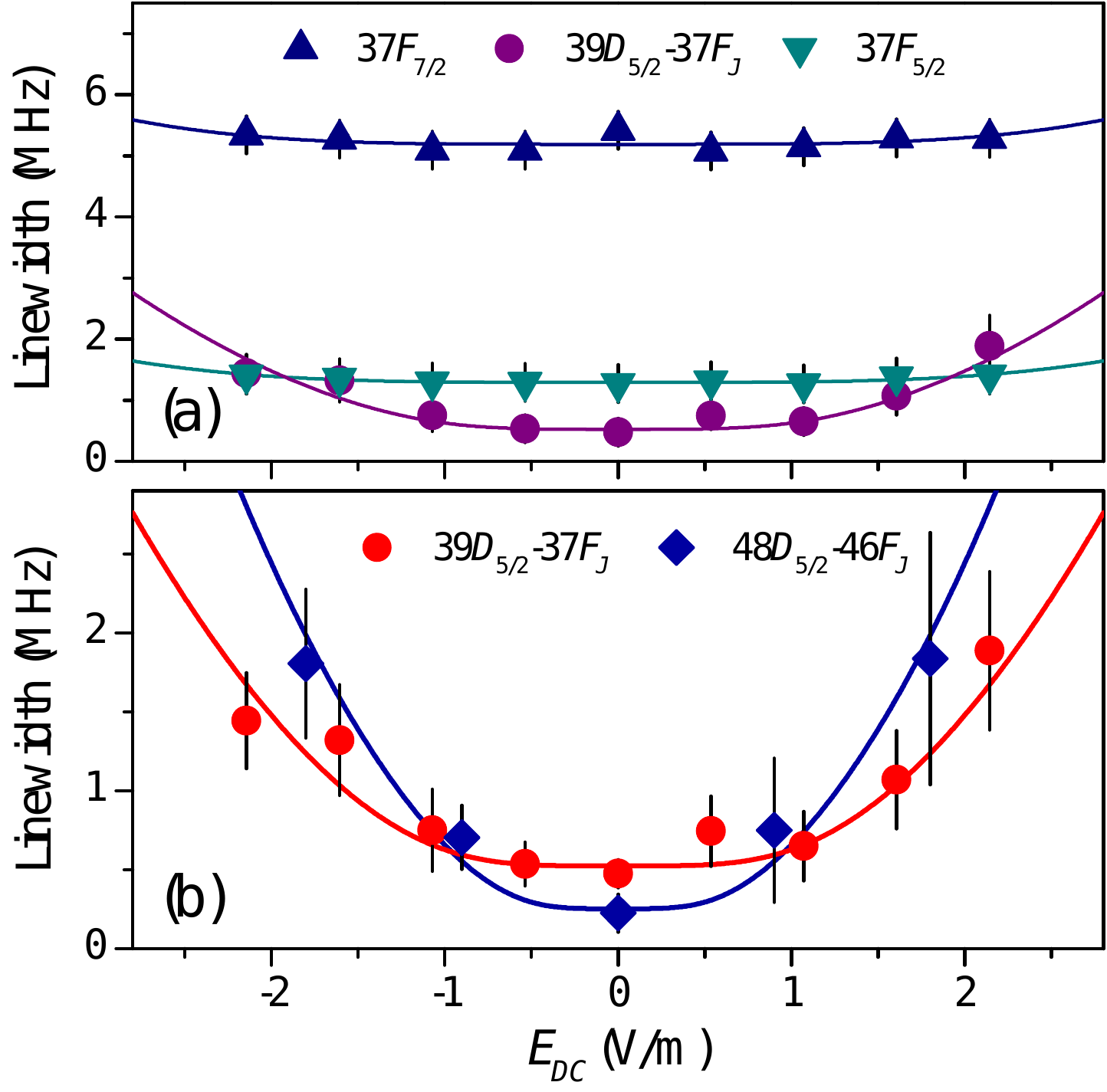}
\end{center}
\caption{(a) Linewidths of the atomic  $37F_{7/2}$ (navy) and $37F_{5/2}$ (dark cyan) peaks, and of the molecular $39D_{5/2}\, 37F_{J}$ (purple) peak from Fig.~\ref{Fig4} vs
$E_{DC}$.
(b) Line broadening of the molecular
$39D_{5/2} \, 37F_{J}$ (red) and $48D_{5/2} \, 46F_{J}$ (royal) molecular peaks vs $E_{DC}$.
In both (a) and (b), solid curves are best fits of the form $a \, \sqrt{1+b \, E_{DC}^{4}}$.
}\label{Fig5}
\end{figure}

In Fig.~\ref{Fig5}(a) we show the linewidths measured for a selection of atomic and molecular peaks as a function of $E_{DC}$.
The atomic lines are fairly broad throughout, in agreement with Fig.~\ref{Fig2}. As a result, any extra broadening that occurs with increasing $|E_{DC}|$ is insignificant.
In contrast, the molecular feature is narrow at $E_{DC} \sim 0$, in agreement with Fig.~\ref{Fig2}, and there is considerable extra broadening with increasing $|E_{DC}|$, considered next.

The variation of the molecular Stark shift is $\delta W_S = - \alpha_M E_{DC} \delta E_{DC} - E_{DC}^2 \delta \alpha_M /2 $. Here, the variation $\delta E_{DC}$ is proportional to  $E_{DC}$ and very small. For the variation of the molecular polarizability we estimate $\delta \alpha_M \sim \pm 0.25 \alpha_M$ (en par with the variation of the relevant atomic tensor polarizabilities). 
Accounting for a background linewidth $\delta W_0$, 
at the largest $|E_{DC}|$ probed in Fig.~\ref{Fig4} the net linewidth $\delta W = \sqrt{\delta W_0^2 + \delta W_S^2}$ would be 
$\delta W \lesssim  |W_S|$. This is not enough to explain the data in Fig.~\ref{Fig5}, where $\delta W \gtrsim 2 |W_S|$ for the molecular peak at the largest $|E_{DC}|$ probed. 

A permanent molecular dipole moment, $p$, could, in principle, explain the large observed $\delta W$. However, the $\gtrsim h \times 10$-MHz strength of Rydberg-pair interactions enforces symmetrized pair states at $E_{DC} = 0$, which have well-defined parity and $p=0$. As such, the situation differs from Rydberg-ground molecules~\cite{Greene2000}, which have extremely weak ``flip'' interactions and allow non-symmetrized pair states with $p \ne 0$. 

We last consider intra-molecular dipole-dipole interactions as a source for the observed large values of $\delta W$. The induced intra-molecular dipoles are parallel to $E_{DC}$ and form an angle $\theta$ relative to the internuclear axis. For constituent polarizabilities estimated at $\alpha_M/2$, the induced dipole-dipole interaction is 
$W_I = E_{DC}^2 (\alpha_M^2/4) (1-3 \cos^2 \theta)/(4 \pi \epsilon_0 R^3)$. With $\theta$ varying between 0 and $\pi$, the variation $\delta W_I = E_{DC}^2 (3 \alpha_M^2)/(16 \pi \epsilon_0 R^3)$. For $39D_{5/2} \, 37F_J$ at $R=2~\mu$m, it is $\delta W_I \gtrsim h \times 100$~kHz. For a given vibrational state, $\delta W_I$ would likely be higher because it would have to be properly averaged over $1/R^3$. Hence, induced dipole-dipole interactions may add to the Stark linewidths and help explain the results in Figs.~\ref{Fig4} and~\ref{Fig5}. Since both the discussed $\delta W_S$ and $\delta W_I$ scale as $E_{DC}^2$, the net linewidths in Fig.~\ref{Fig5} should follow a function $\delta W = a \sqrt{1+ b \, E_{DC}^4}$. This function, applied in Fig.~\ref{Fig5}, fits the data indeed quite well. 


In summary, we have observed FS-induced $(n+2)D_{5/2} \, nF_J$ Cs Rydberg macro-dimers, modeled the data, and explained the formation mechanisms of these molecules. Different from the commonly used method of laser photo-association, we have employed microwave photo-association, which is narrow-band and sensitive to van-der-Waals interactions of the launch Rydberg-pair states. In the future, improved atom cooling methods, lattice-trapped Rydberg atoms, and Franck-Condon tuning principles could be used to reduce the linewidths and to expand the utility of microwave photo-association spectroscopy. The research may reveal detailed vibrational structures suitable to test and improve models for Born-Oppenheimer potentials, including the effect of non-adiabatic couplings on vibrational-level energies and lifetimes. Future research into the mixing of Rydberg macro-dimers with high-$\ell$ levels may shed light onto unresolved issues regarding our measured molecular polarizabilities. Intra-molecular induced-dipole-dipole interactions in Rydberg macro-dimers, for which we have found indications, may also deserve additional work.

We gratefully acknowledge the National Natural Science Foundation of China (Grants No. 12120101004,  No. 12241408, No. 61835007, and No. 62175136); the Changjiang Scholars and Innovative Research Team in Universities of the Ministry of Education of China (Grant No. IRT 17R70); and the 1331 project of Shanxi Province. G.R. acknowledges support by the University of Michigan.

\bibliography{reference}

\section{Supplement}

\subsection{Adiabatic potentials}

To model the experimental spectra, we consider the product Hilbert space of two Rydberg atoms, and add the multipole interaction between the atoms to the unperturbed two-body Hamiltonian. The molecular potential energy curves are obtained by diagonalization of the Rydberg-pair Hamiltonian within basis sub-sets of states with conserved $M$, the sum of the projections of the total electronic angular momenta onto the internuclear axis, and conserved parity (which depends on whether the sum of the orbital electronic angular momenta is even or odd). We include multipolar interactions up to order $L_a + L_b $=6, with the dipole, quadrupole etc. terms for atoms ``a'' and ``b'' denoted $L_{a}=1, 2,... $ and $L_b=1, 2,... $. The multipole interactions scale as $n^{2(L_a+L_b)}/R^{1+L_a+L_b}$, where $R$ is the internuclear distance and $n$ the average principal quantum number of the Rydberg-pair states of interest. The leading interaction of a pair of atoms in states $a=(n+2)D_{5/2}$ and $b=nF_J$, which are of main interest in the present work, is the dipole-dipole interaction, which scales as $n^4/R^3$. The interactions drop rapidly in importance with increasing order $L_a+L_b$. In our calculations we are setting the cutoff at $L_a + L_b \le 6$, which is generous. More details describing our calculations are found in References~\cite{Han2018,Han2019} in the main text.

\subsection{Excitation rates}

To compute model spectra, we consider the microwave excitation rates from the optically-prepared launch states $[(n+2)D_{5/2}]_2$ into pair states $(n+2)D_{5/2} \, nF_J$. At the $R$-values of the $DF$-macro-dimers of interest, molecular-potential calculations for the launch $[(n+2)D_{5/2}]_2$ pair states show positive van-der-Waals level shifts of $\lesssim 2$~MHz for $39D$. Hence, within a reasonable approximation, we can assume that the $[(n+2)D_{5/2}]_2$ pair states are only weakly perturbed.
We therefore compute the microwave excitation rates between non-interacting
$[(n+2)D_{5/2}]_2$ atom pairs in the microwave transition's ground state, and strongly-perturbed and -mixed
$(n+2)D_{5/2} \, nF_J$ atom pairs in the microwave transition's excited state. This allows
us to use the same methods as described
in References~\cite{Han2018,Han2019} in the main text.
The rates, averaged over the angular alignment of the atom pairs relative to the microwave field, are shown by the areas of the dots in Fig.~\ref{Fig2}~(a) in the main text. 
To model the experimental spectra in a two-body approximation, in which interactions between more than two atoms are ignored, we then compute radially-averaged model spectra in which we sum over all $M$-values and parities, and we apply a radial weighting factor $(3/r) (R/r)^2 \exp[-(R/r)^3]$  (the nearest-neighbor distribution in an ideal gas, with $r$ denoting the Wigner-Seitz radius). Typical results are shown in Fig.~\ref{Fig2}~(c) in the main text for two indicated values of $r$.

\subsection{Atomic electric polarizabilities}

We first calculate the polarizabilities of the atomic constituents (using private codes). For
the $39D_{5/2} \, 37F_J$ macro-dimer, these are
$a_{s}(39D_{5/2})=-11.7$~kHz/(V/m)$^2$, $a_{t}(39D_{5/2})= 13.0$~kHz/(V/m)$^2$,
$a_{s}(37F_{5/2})= 75.3$~kHz/(V/m)$^2$, $a_{t}(37F_{5/2})=-26.4$~kHz/(V/m)$^2$,
$a_{s}(37F_{7/2})= 74.7$~kHz/(V/m)$^2$, and $a_{t}(37F_{7/2})=-30.5$~kHz/(V/m)$^2$, where
the subscripts
$s$ and $t$ stand for scalar and tensor polarizabilities. We recite that for a given $m_J$, with the direction of $E_{DC}$ taken as the quantization axis, the polarizability is $\alpha(\lambda) = a_{s,\lambda} + a_{t,\lambda} (3 m_J^2 - J(J+1))/J/(2J-1))$, and the level shift is $s(\lambda) = -(1/2) \alpha(\lambda) E_{DC}^2$, where $\vert \lambda \rangle$ is a short-hand for the quantum state $ \vert n, \ell, J, m_J \rangle$ in the $E_{DC}$-frame. Taking an isotropic average, both $37F_J$ lines in Fig.~\ref{Fig4} in the main text are expected to shift according to the average differential polarizability between the upper $37F_J$ and the lower $39D_{5/2}$ states. This average computes to be $\approx 85$~kHz/(V/m)$^2$, which agrees within uncertainties with the values derived from the Stark-shift fits in Fig.~\ref{Fig4} of the main text. For completeness we recall here that isotropic averages of the tensor polarizabilities are zero.

\subsection{Molecular electric polarizabilities}

The symmetrized state of a Rydberg pair is of the form
\begin{eqnarray}
 \vert \psi_{\rm {mol}} \rangle &=& \frac{1}{\sqrt{2}} \sum_{\lambda_a \lambda_b} c_{ab}
(\vert \lambda_a \lambda_b \rangle \nonumber -p (-1)^{\ell_a + \ell_b} \vert \lambda_b  \lambda_a \rangle)
\end{eqnarray}
with $\lambda_{a(b)}=\vert n_{a(b)}, \ell_{a(b)}, J_{a(b)}, m_{J_{a(b)}} \rangle$ denoting single-atom states
in the molecular frame, and $a$ and $b$ being subscripts for the two atoms. The coefficients $c_{ab} (R)$ are obtained in molecular-potential calculations explained above. Rotating the state from the molecular into the $E_{DC}$-frame with (reduced) rotation operators $d^{(J)}_{\tilde{m_J}, m_J} (\theta)$~
[see, {\sl{e. g.}}, J. J. Sakurai and J. Napolitano, ``Modern Quantum Mechanics'', 
Cambridge University Press (2020)] with $\theta$ denoting the angle between $E_{DC}$ and the internuclear axis, the molecular state in the $E_{DC}$-frame takes the form
\begin{eqnarray}
 \vert \psi_{\rm {mol}} \rangle &=& \frac{1}{\sqrt{2}} \sum_{\tilde{\lambda}_a \tilde{\lambda}_b} \tilde{c}_{ab}
(\vert \tilde{\lambda}_a \tilde{\lambda}_b \rangle \nonumber -p (-1)^{\ell_a + \ell_b} \vert \tilde{\lambda}_b  \tilde{\lambda}_a \rangle),
\end{eqnarray}
with the transformed coefficients
\[\tilde{c}_{ab} = c_{ab}
d^{(J_a)}_{\tilde{m}_{Ja}, m_{Ja}} d^{(J_b)}_{\tilde{m}_{Jb}, m_{Jb}} \quad.
\]
In a perturbative picture, which should apply as long as the molecular Stark shift is much less than the binding energy (a few MHz in our case), the leading contribution of
the molecular polarizability is the sum of the polarizabilities of the atomic constituents. This yields a theoretical molecular polarizability of
\begin{equation}
\alpha_{M,T} = \sum_{\tilde{\lambda}_a \tilde{\lambda}_b} \vert \tilde{c}_{ab} \vert^2 [\alpha(\tilde{\lambda}_a) +
\alpha(\tilde{\lambda}_b)] \quad.
\end{equation}
The result is averaged over $\theta$ with the normalized weighting function $\sin(\theta)/2$ with $\theta$ ranging from 0 to $\pi$.

We have performed the calculation for electronic adiabatic states in the potential wells in Fig.~\ref{Fig2}~(a) of the main text and found $\alpha_{M,T}$-values close to the sum of the $39D_{5/2}$ and $37F_J$ scalar atomic polarizabilities. These are also the atomic states relevant to Fig.~\ref{Fig4} of the main text.
Noting that the adiabatic states in the potential wells in Fig.~\ref{Fig2}~(a) of the main text carry only very small contributions from states with $\ell \ne 2, 3$, for the basis sets used, this result is not very surprising. However,
$\alpha_{M,T}$ amounts to only $\approx 1/2.7$ of the measured value, $\alpha_M$.

\end{document}